# Nanoscale optical imaging of multi-junction MoS$_2$-WS$_2$ lateral heterostructure

Jiru Liu[1], Wenjin Xue[1], Haonan Zong[1], Xiaoyi Lai[1], Prasana K. Sahoo[2], Humberto R. Gutierrez[2] and Dmitri V. Voronine[2]

[1]University of Science and Technology of China, Hefei, 230026, China
[2]Department of Physics, University of South Florida, Tampa, FL 33620, USA

## Abstract

Two-dimensional monolayer transition metal dichalcogenides (TMDs) have unique optical and electronic properties for applications pertaining to field effect transistors, light emitting diodes, photodetectors, and solar cells. Vertical interfacing of WS$_2$ and MoS$_2$ layered materials in combination with other families of 2D materials were previously reported. On the other hand, lateral heterostructures are particularly promising for the spatial confinement of charged carriers, excitons and phonons within an atomically-thin layer. In the lateral geometry, the quality of the interface in terms of the crystallinity and optical properties is of paramount importance. Using plasmonic near-field tip-enhanced technology, we investigated the detailed nanoscale photoluminescence (nano-PL) characteristics of the hetero-interface in a monolayer WS$_2$-MoS$_2$ lateral heterostructure. Focusing the laser excitation spot at the apex of a plasmonic tip improved the PL spatial resolution by an order of magnitude compared to the conventional far-field PL. Nano-PL spatial line profiles were found to be more pronounced and enhanced at the interfaces. By analyzing the spectral signals of the heterojunctions, we obtained a better understanding of these direct band gap layered semiconductors, which may help to design next-generation smart optoelectronic devices.

**Introduction**

Two-dimensional (2D) monolayer transition metal dichalcogenides (TMDs) have attracted a wide attention due to their layer dependent band gap properties[1,2]. Several proof-of-concept devices were demonstrated such as FET, photodiode, LED and solar cell. Most commonly used 2D TMD materials are $WS_2$ and $MoS_2$ monolayers in which the transition metal layer of W or Mo covalently binds with two chalcogenide (S) layers in the form of the S-X-S pattern[3,4] (Fig. 1). Given the promising potential applications in optoelectronic devices in the form of the p-n junction[5-7], scalable production of lateral heterostructures is quite a challenging task[4,8]. Nevertheless, atomically sharp multi-junctions in lateral heterostructures of monolayer $WS_2$ and $MoS_2$ were recently reported using a one-pot synthesis process (Sahoo, et al). It is possible to directly fabricate multi-stripes of different TMD materials by changing the reactive gas environment in the presence of the water vapor in a chemical vapor deposition (CVD) system[4]. In this report, we investigated the nanoscale optical properties of a CVD-grown multi-junction monolayer $WS_2$-$MoS_2$ lateral heterostructure on $SiO_2$ substrate[4].

**Experiment**

In order to understand the nanoscale photoluminescence (PL) properties at the heterojunction, we used a plasmonic gold-coated silver tip of a scanning probe microscope coupled to a confocal optical microscope (Horiba) to enhance the PL signal from the boundary of the two materials.

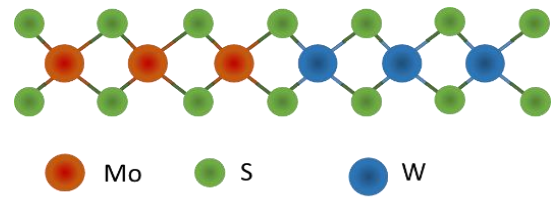

**Figure 1**. Schematic representation of monolayer $MoS_2$-$WS_2$ lateral heterostructure.

By focusing the laser on the plasmonic tip, the PL intensity was enhanced by ~30 times[9]. The size of the tip (~ 10 – 20 nm radius) determines the size of excitation area and determines the spatial resolution. The tip-sample distance was controlled with sub-nanometer precision and optical images were recorded with the tip close to the sample (near-field, NF) and with the tip far from the sample (far-field, FF)[10-12].

We used two different laser excitation wavelengths, λ = 660 nm (red) and 532 nm (green). Monolayer $WS_2$ and $MoS_2$ have different exciton energy gaps[4]. Hence, the PL spectra provide a unique identification of the type of the material used as well as the nature of the interface[13]. For monolayer $WS_2$, we observed the PL peak position is at ~ 630 nm (1.97 eV), and for $MoS_2$ at ~ 674 nm (1.84 eV)[4,14,15]. 660 nm laser (1.88 eV) could only excite $MoS_2$ while the 532 nm laser (2.33 eV) could excite both materials. While using the 532 nm laser excitation, it was found that the signal from the monolayer $WS_2$ domain was very strong compared to $MoS_2$. In fact, the PL spectra were mostly dominated by $WS_2$. Thus, the two laser excitation sources used in this study were selectively used to excite the PL spectra of the individual $MoS_2$ and $WS_2$ domains.

Thin multiple stripes were observed in the $WS_2$-$MoS_2$ lateral heterostructure that correspond to heterojunctions. Coincidentally, two bright PL lines which are closely adjacent to each other were found in scanning map of the near field PL intensity. Analysis to these PL lines revealed that their positions overlap with the boundaries between the $WS_2$-$MoS_2$; strongly suggesting that the bright PL lines mostly originate from the heterojunction regions.

**Results and discussion**

Atomic force microscopy (AFM) was used to obtain the topographic and phase information of the lateral heterostructure (Fig. 2), prior to the tip-enhanced photoluminescence (TEPL) maps. Figure 2a shows the topography of a multi-junction monolayer $WS_2$-$MoS_2$ lateral heterostructure, indicating uniformity across the triangular domain. The corresponding phase map shows the existence of multiple thin strips of $MoS_2$ and $WS_2$ monolayers (Fig. 2b). As mentioned above, for the TEPL maps, we focused the 660 nm laser onto a gold-coated silver tip, and scanned a part of the triangle (inset in Fig. 2b). Bright lines were clearly observed in the NF image, but the stripes of $MoS_2$ were so thin that it was hard to distinguish the two lines in the FF. In Fig. 2b, specific areas were chosen to perform imaging of the two lines with a high spatial resolution. Two bright lines were clearly observed in the NF.

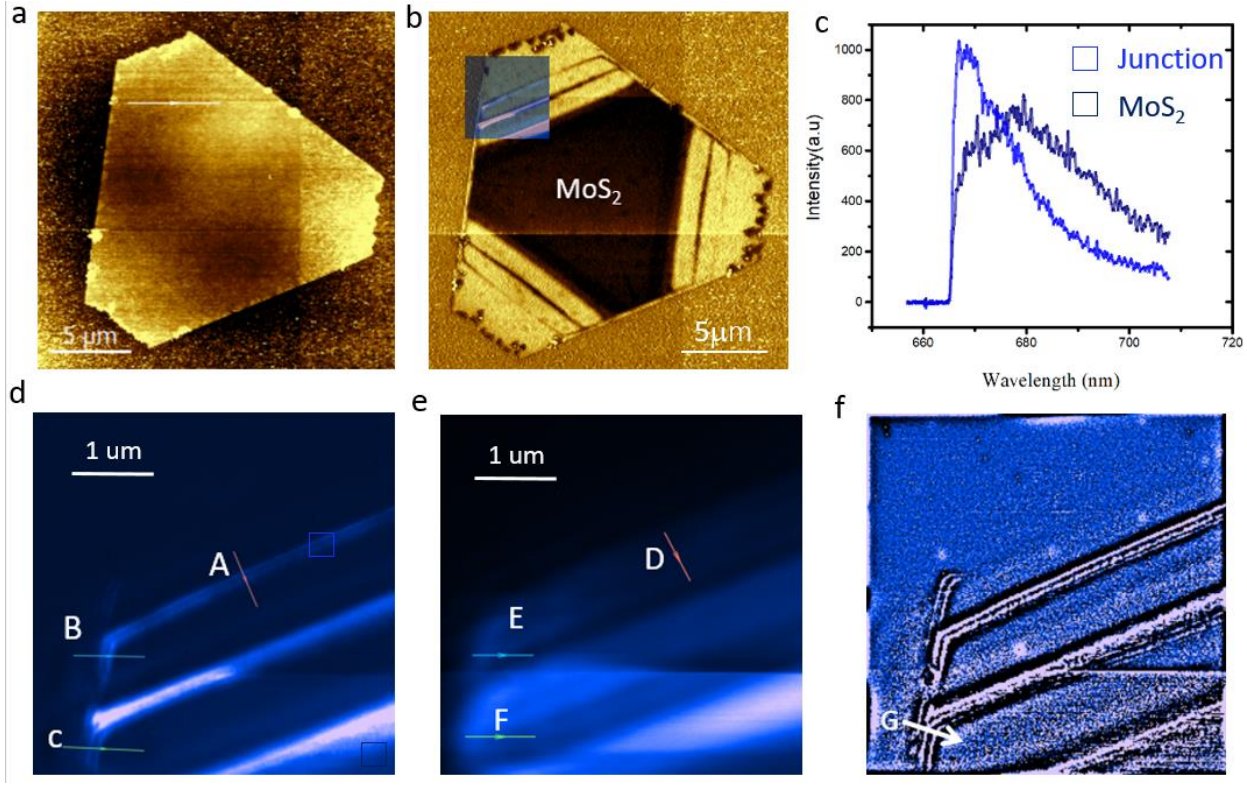

**Figure 2.** Nanoscale imaging of the multi-junction $MoS_2$-$WS_2$ lateral heterostructure. **a.** AFM topography image and **b** AFM phase image; inset shows the TEPL scanning area. **c.** Nano-PL spectra collected from the box region of near filed image (d). **d.** NF PL map and **e.** FF PL map of the selected area in (b). **f.** Fourier filtered image of the selected area in (b) clearly shows bright PL lines.

Fourier processing was used to remove the background noise. After processing, the bright lines become clearer in Fig. 2f. The sulfide based TMD lateral heterostructures have truncated triangular geometry. The $WS_2$ has a different growth rate along the two prominent edges of the initial $MoS_2$ domain. This gives rise to the anisotropic growth of the $WS_2$ domain along different directions in this kind of heterostructure. There exist very narrow lines in one of the edges which is very difficult to observe in the FF optical microscope images. However, these narrow strips were clearly resolved in the NF images (Figs. 2f and 3f). The number of PL lines matched with the number of the heterojunctions. We further evaluated the presence of different material domains and heterojunctions using the peak position of the PL spectra in the TEPL maps. The nano-PL peak position associated with the direct excitonic emission for monolayer $WS_2$ was observed at ~ 630 nm. For $MoS_2$ the direct excitonic emission was recorded at ~ 674 nm. This further supports the presence of distinct crystalline domains in the lateral heterostructure.

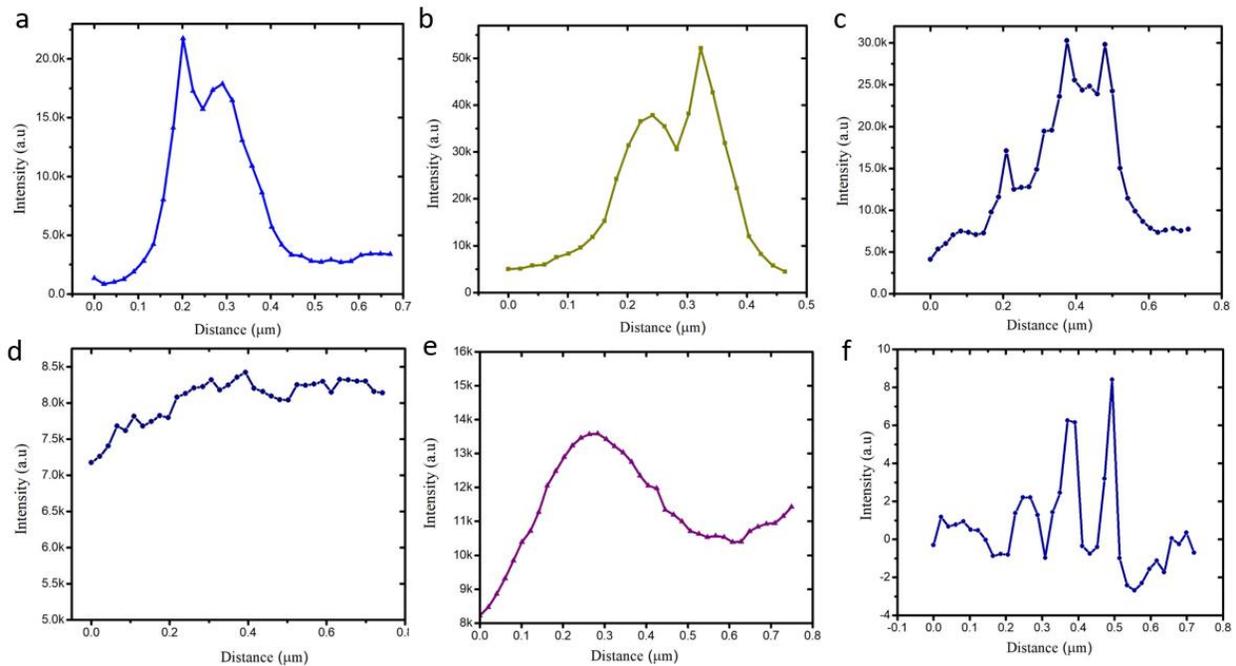

**Figure 3.** Nano-PL intensity line profiles from the PL maps with 660 nm laser excitation shown in Fig. 2. **a-c.** NF PL intensity line profiles indicated by A, B and C, respectively, in Fig. 2d. **d** and **e** show the FF PL intensity line profiles indicated by white lines D and E in Fig. 2e. **f.** Fourier filtered NF PL intensity line profile indicated by the white line G in Fig. 2e.

The NF PL line profiles in Figs. 3a – 3c across different junction regions in the PL maps (indicated in Figs. 2d and e) show that the $MoS_2$ and $WSe_2$ domains are distinguishable and laterally well separated. The corresponding FF PL signals fail to resolve the junctions. Furthermore, the nano-PL intensity line profile at the cross-section of the NF map within the triangular domain (Fig. 3f) shows a more prominent variation of the PL intensity. The sharpness of the hetero-interface further depends on the transition from $MoS_2{\rightarrow}WS_2$ (diffuse interface) or $WS_2{\rightarrow}MoS_2$ (sharp interface) domains. These characteristics of different hetero-interfaces depend on the carrier gas switching cycle during the sequential growth of the individual TMD domains in the CVD system.

**Conclusions**

Tip-enhanced near-field optical microscopy provides a route to explore nanoscale optical properties with sub-diffraction resolution of CVD-grown multi-junction monolayer $MoS_2$-$WS_2$ lateral heterostructures. Nano-PL signals were significantly enhanced at the hetero-interfaces. We demonstrated that NF imaging provides a better nanoscale spatial resolution compared to the FF mode. Thus, NF nano-PL maps can be used to probe nanoscale material heterogeneity in a wide range of 2D heterostructures. This technique can be further used as an effective optical characterization tool for the development of smart 2D optoelectronic devices with a high quantum yield.

**Acknowledgements**

We thank Marlan Scully, Alexei Sokolov and Zhe He for helpful discussions. DVV acknowledges the support by the National Science Foundation (grant number CHE-1609608). H.R.G. acknowledged support by the National Science Foundation Grant DMR-1557434.